# Similarity-and-Independence-Aware Beamformer with Iterative Casting and Boost Start for Target Source Extraction Using Reference

Atsuo Hiroe, *Member, IEEE*

**Abstract** Target source extraction is significant for improving human speech intelligibility and the speech recognition performance of computers. This study describes a method for target source extraction, called the similarity-and-independence-aware beamformer (SIBF). The SIBF extracts the target source using a rough magnitude spectrogram as the reference signal. The advantage of the SIBF is that it can obtain a more accurate signal than the spectrogram generated by target-enhancing methods such as speech enhancement based on deep neural networks. For the extraction, we extend the framework of deflationary independent component analysis (ICA) by considering the similarities between the reference and extracted target sources, in addition to the mutual independence of all the potential sources. To solve the extraction problem by maximum-likelihood estimation, we introduce three source models that can reflect the similarities. The major contributions of this study are as follows. First, the extraction performance is improved using two methods, namely boost start for faster convergence and iterative casting for generating a more accurate reference. The effectiveness of these methods is verified through experiments using the CHiME3 dataset. Second, a concept of a fixed point pertaining to accuracy is developed. This concept facilitates understanding the relationship between the reference and SIBF output in terms of accuracy. Third, a unified formulation of the SIBF and mask-based beamformer is realized to apply the expertise of conventional BFs to the SIBF. The findings of this study can also improve the performance of the SIBF and promote research on ICA and conventional beamformers.

## I. Introduction

The process of extracting the target source from mixtures of multiple sound sources, such as denoising and speech extraction, plays a significant role in improving the speech intelligibility of humans and the automatic speech recognition (ASR) performance of computers [1]. The associated methods are generally classified into nonlinear and linear methods. In the last decade, nonlinear methods have significantly improved due to the development of deep neural networks (DNNs). These methods, referred to as DNN-based speech enhancements (SEs), can generate cleaner speech from a noisy speech [2][3] and extract an utterance from overlapping speeches of multiple speakers [4][5]. Meanwhile, linear methods such as the beamformer (BF) are advantageous in avoiding nonlinear distortions such as musical noises and spectral distortions [6][7].

To incorporate the features of linear and nonlinear methods, we developed a novel BF referred to as a *similarity-and-independence-aware beamformer* (SIBF) in a previous study [8]. The SIBF uses the magnitude spectrogram generated by any target-enhancing method (including DNN-based SE) as a reference for the target source. Unlike conventional BFs, which require one or more time-frequency (TF) masks generated by the DNN [9][10][11][12][13], the SIBF can be combined with various types of DNNs that output a waveform and TF mask, as well as with a spectrogram, because the data can be converted into a magnitude spectrogram.

In the framework of the SIBF, its output is required to be more accurate than the given reference because, if this does not hold, the reference should be used directly. To satisfy this requirement, the SIBF is formulated as follows. The framework of the deflationary independent component analysis (ICA) [14] is extended to utilize the dependence between the reference and the estimated target source, in addition to the independence of all the sources. This characteristic enables the SIBF to generate an estimated target source more accurately than the reference. To represent the dependence, we examined two source models, namely the time-frequency-varying variance (TV) Gaussian model and bivariate spherical (BS) Laplacian model. As a reference, the SIBF requires a magnitude spectrogram corresponding to only the target source, unlike the conventional reference-based ICA using magnitude spectrograms [15][16], which requires references corresponding to all sources, even if the number of sources is unknown.

In this study, we realize the following advancements for the SIBF.

| Signal name | Spectrogram | An element | Column vector of all channel elements |
|---|---|---|---|
| **Source** | $\mathbf{S}_k \in \mathbb{C}^{F \times T}$ | $s_k(f,t) \in \mathbb{C}$ | $\mathbf{s}(f,t) = [s_1(f,t), ..., s_M(f,t)]^{\mathrm{T}}$ |
| **Observation** | $\mathbf{X}_k \in \mathbb{C}^{F \times T}$ | $x_k(f,t) \in \mathbb{C}$ | $\mathbf{x}(f,t) = [x_1(f,t), ..., x_N(f,t)]^{\mathrm{T}}$ |
| **Uncorrelated observation** | $\mathbf{U}_k \in \mathbb{C}^{F \times T}$ | $u_k(f,t) \in \mathbb{C}$ | $\mathbf{u}(f,t) = [u_1(f,t), ..., u_N(f,t)]^{\mathrm{T}}$ |
| **Estimated source** | $\mathbf{Y}_k \in \mathbb{C}^{F \times T}$ | $y_k(f,t) \in \mathbb{C}$ | $\mathbf{y}(f,t) = [y_1(f,t), ..., y_N(f,t)]^{\mathrm{T}}$ |
| **Reference** | $\mathbf{R} \in \mathbb{R}^{F \times T}$ | $r(f,t) \in \mathbb{R}$ | (not available) |

The first advancement is the improvement of the extraction performance. As discussed in the previous study, the SIBF has a minimum of two options for improvement: the examination of model-specific properties, and the generation of a more accurate reference. For the model-specific properties in this study, we examine (i) the TV Student's t model as an alternative source model and (ii) a technique for faster convergence, referred to as *boost start*. To generate a more accurate reference, we examine a technique called *iterative casting*, which casts the output of the SIBF into the DNN for reference generation. The second advancement is the development of a novel concept of a fixed point pertaining to accuracy. This concept is conceived by discussing the results of iterative casting, and it facilitates understanding the various behaviors of the SIBF. The third advancement is the unified formulation of the SIBF and mask-based BF, e.g., the maximum signal-to-noise ratio (Max SNR) BF [9][10]. This enables us to apply expertise on the mask-based BF to the SIBF.

The remainder of this paper is organized as follows. Section II presents the formulation of the SIBF problem and derives the rules required to obtain the filter for the extraction of the target source. Section III presents an investigation of the techniques used for improving the extraction performance, namely the boost start and iterative casting, mentioned previously as the first advancement. Section IV details a series of experiments conducted to determine the optimal hyperparameters for each source model, to confirm the effectiveness of boost start and iterative casting, and to compare the SIBF with the conventional mask-based BFs. Section V discusses the various behaviors of the iterative casting by introducing the concept of the fixed point, mentioned previously as the second advancement. Section VI clarifies the relationship between the SIBF and two categories of related works, namely reference-based ICA and mask-based Max SNR BF, mentioned previously as the third advancement. Finally, Section VII presents the conclusions of this study.

## II. PROBLEM FORMULATION

The notations listed in Table I are used consistently throughout this paper to represent the TF domain signals, with $f$, $t$, and $k$ denoting the indices of the frequency bin, frame, and channel, respectively. Each spectrogram consists of $F$ frequency bins and $T$ frames.

Fig. 1 shows the workflow of the proposed SIBF. The inputs are the multichannel observation spectrograms obtained from multiple microphones, and the output is a spectrogram of the target source. A magnitude spectrogram corresponding to the target source is used as the reference and can be estimated using various methods, including the DNN-based SE. Such a reference should be considered as "rough" (or less accurate) for the following:

1. It includes several nondominant interferences (sources other than the target source) or can be distorted due to the excessive removal of these interferences.
2. It does not contain any phase information.

The workflow involves two steps: (i) estimating the rough magnitude spectrogram of the target source, and (ii) applying the SIBF with the rough spectrogram as the reference.

To realize this process, we extend the framework of the deflationary ICA using decorrelation (pre-whitening) [14], as presented in the following subsections.

### A. Mixing and Separating Processes

Fig. 2 presents the modeling procedure of the mixing and separating processes of the SIBF, where the dashed box represents the extension for the SIBF, and the remaining parts are based on the conventional deflationary ICA.

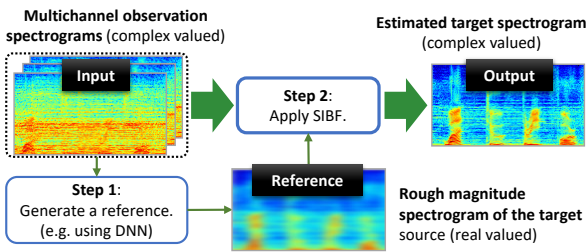

Fig. 1: Workflow of SIBF

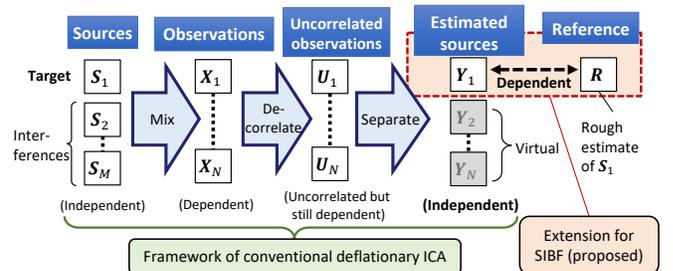

Fig. 2: Modeling procedure of the mixing and separating processes. The unique points of the modeling are as follows: (i) the consideration of the dependence between $\mathbf{Y}_1$ and $\mathbf{R}$, in addition to the independence, and (ii) the estimation of only $\mathbf{Y}_1$, whereas $\mathbf{Y}_2, ..., \mathbf{Y}_N$ are virtual estimated sources.

First, let us explain the conventional parts. Sources $S_1, \dots, S_M$ are assumed to be mutually independent. Without loss of generality, $S_1$ is considered the target source, namely the source of interest in this study, whereas the other sources are considered as interferences. The observations $X_1, \dots, X_N$ represent the spectrograms obtained from $N$ microphones. In the TF domain, each observation spectrogram $X_k$ is approximated as an instantaneous mixture of the sources.

To apply the framework of deflationary ICA, we generate uncorrelated observations $U_1, \dots, U_N$. Using the decorrelation matrix $P(f)$, the decorrelation process in each frequency bin is expressed as follows:

$$u(f,t) = P(f)x(f,t), \tag{1}$$
$$\text{s.t. } \langle u(f,t)u(f,t)^{\mathrm{H}} \rangle_t = I, \tag{2}$$

where $\langle \cdot \rangle_t$ and $I$ denote the average over $t$ and the identity matrix, respectively. The matrix $P(f)$ is computed as follows:

$$\Phi_x(f) = \langle x(f,t)x(f,t)^{\mathrm{H}} \rangle_t, \tag{3}$$
$$\Phi_x(f) = Q(f)\Lambda(f)Q(f)^{\mathrm{H}}, \tag{4}$$
$$P(f) = \Lambda(f)^{-1/2}Q(f)^{\mathrm{H}}, \tag{5}$$

where $\Phi_x(f)$, $Q(f)$, and $\Lambda(f)$ denote a covariance matrix of the observation spectrograms, matrix consisting of all the eigenvectors of $\Phi_x(f)$, and diagonal matrix consisting of all the corresponding eigenvalues, respectively.

From the assumption of the sources, the estimated sources $Y_1, \dots, Y_N$ are also mutually independent. The separation process is expressed by (6), using a separation matrix $W(f)$ that makes $y_1(f,t)$, ..., $y_N(f,t)$ mutually independent.

$$y(f,t) = W(f)u(f,t) = W(f)P(f)x(f,t). \tag{6}$$

Considering the decorrelation, we can restrict $W(f)$ to a unitary matrix, such that $W(f)W(f)^{\mathrm{H}} = I$ [14].

In conventional ICA, it is uncertain which source corresponds to the first estimated source $Y_1$. To solve this issue, or in other words, to certainly associate $Y_1$ with the target source $S_1$, we extend this framework. In the extension shown in Fig. 2, the reference $R$ is a rough estimate of $S_1$, as mentioned in the beginning of this section. We consider the dependence between $Y_1$ and $R$ in addition to the independence of all the estimated sources. The objectives are to make $Y_1$ both similar to the reference (using the dependence) and more accurate than the reference (using the independence).

In Fig. 2, the estimated sources other than $Y_1$ are unnecessary because $S_1$ is the only source of interest. Thus, to generate only $Y_1$, we employ the deflationary estimation [14], i.e., one-by-one separation. This indicates that the other estimated sources, $Y_2, \dots, Y_N$, are virtual (or potential).

To extract only the estimated target source $y_1(f,t)$, we use (7) instead of (6).

$$y_1(f,t) = w_1(f)^{\mathrm{H}}u(f,t) = w_1(f)^{\mathrm{H}}P(f)x(f,t), \tag{7}$$

where $w_1(f)^{\mathrm{H}}$ is the first-row vector in $W(f)$, thus satisfying $w_1(f)^{\mathrm{H}}w_1(f) = 1$. We refer to $w_1(f)$ as an *extraction filter*. Unlike that in our previous study [8], $w_1(f)$ itself is a column vector. Based on this constraint and the effect of the decorrelation, $y_1(f,t)$ is constrained such that $\langle |y_1(f,t)|^2 \rangle_t = 1$. Because of this constraint, we must estimate a suitable scale of the estimated target source after obtaining the filter, as mentioned in Section II.D.

### B. Maximum-likelihood Estimation of Target Source

We can solve the target source extraction problem shown in Fig. 2 by maximum-likelihood (ML) estimation [17], which is extensively used in blind source separation (BSS) problems such as the ICA. In the subsequent explanation, p($A$) and p($A \mid B$) denote the probability density function (PDF) of $A$ and the conditional PDF of $A$ given $B$, respectively. If $A$ consists of multiple arguments or elements, these functions represent joint PDFs.

Let $\mathcal{W} = \{W(1), \dots, W(F)\}$ be a set of separation matrices for all frequency bins; this is a parameter set to be estimated. For the objective function that represents both the independence of $Y_1, \dots, Y_N$ and the dependence between $Y_1$ and $R$, as shown in Fig. 2, we consider the temporally averaged negative log-likelihood $L(\mathcal{W})$ of the reference and observations:

$$L(\mathcal{W}) = -\frac{1}{T}\log \mathrm{p}(R, X_1, \cdots, X_N | \mathcal{W}). \tag{8}$$

To derive the rule for the extraction filter, we modify (6) as follows:

$$\begin{bmatrix} r(f,t) \\ y(f,t) \end{bmatrix} = \begin{bmatrix} 1 & 0 \\ 0 & W(f) \end{bmatrix}\begin{bmatrix} 1 & 0 \\ 0 & P(f) \end{bmatrix}\begin{bmatrix} r(f,t) \\ x(f,t) \end{bmatrix} \tag{9}$$
$$\Leftrightarrow y'(f,t) = W'(f,t)P'(f,t)x'(f,t). \tag{10}$$

For simplicity, we assume that the number of sources is equal to the number of microphones, that is, $M = N$, and all elements in the same spectrogram are mutually independent. From these assumptions, we can express (8) as (11).

$$L(\mathcal{W}) = -\frac{1}{T}\sum_t\sum_f\log \mathrm{p}(\underbrace{r(f,t), x_1(f,t), \cdots, x_N(f,t)}_{x'(f)} | W(f)) \tag{11}$$

In the formulation of ICA, the likelihood of the observations can be expressed by using both the estimated sources and separation matrix [18]. Similarly, we can express (11) as (12), regarding $y'(f,t)$ and $W'(f,t)P'(f,t)$ in (10) as the estimated sources and separation matrix, respectively.

$$L(\mathcal{W}) = -\sum_f \langle \log p(\underbrace{r(f,t), y_1(f,t), \cdots, y_N(f,t)}_{\boldsymbol{y}'(f)}) \rangle_t \\ - 2\sum_f \log|\det(\boldsymbol{W}'(f)\boldsymbol{P}'(f))|. \quad (12)$$

The last term in (12) is constant, considering that (13) is satisfied if $\boldsymbol{W}(f)$ is a unitary matrix.

$$\det(\boldsymbol{W}'(f)\boldsymbol{P}'(f)) = \det(\boldsymbol{W}(f)\boldsymbol{P}(f)) \\ = \det(\boldsymbol{P}(f)). \quad (13)$$

We can express (12) as (14), because the dependence is only retained between $r(f,t)$ and $y_1(f,t)$.

$$L(\mathcal{W}) = -\sum_f \langle \log p(r(f,t), y_1(f,t)) \rangle_t \\ - \sum_f \sum_{k \geq 2} \langle \log p(y_k(f,t)) \rangle_t + \text{const.} \quad (14)$$

We refer to $p(r(f,t), y_1(f,t))$ as *the source model*, which is detailed in the following subsection.

By minimizing (14), we can estimate the most likely sources. To estimate $\boldsymbol{w}_1(f)$, which is the extraction filter for $y_1(f,t)$, we minimize only the first term on the right-hand side of (14) with respect to $\boldsymbol{w}_1(f)$, as follows:

$$\boldsymbol{w}_1(f) = \arg\min_{\boldsymbol{w}_1(f)} \{-\langle \log p(r(f,t), y_1(f,t)) \rangle_t\} \quad (15)$$

$$\text{s.t. } \boldsymbol{w}_1(f)^{\mathrm{H}} \boldsymbol{w}_1(f) = 1. \quad (16)$$

It should be noted that $r(f,t)$ is treated as constant in this minimization problem, whereas $y_1(f,t)$ is to be optimized depending on $\boldsymbol{w}_1(f)$. This is because the reference has been generated employing a target-enhancing method.

### C. Specific Source Models and Corresponding Rules for Extraction Filter

In the formulation of the SIBF, the source models represent the distribution of the target source and the dependence between the estimated target source and reference. In this study, we examine three models:
1. TV Gaussian model
2. BS Laplacian model
3. Time-frequency-varying variance complex Student's t (TV t) model

Each source model is explained, and the corresponding rules for the extraction filter are derived, which are referred to as *update rules*. As presented in the following subsections, the scale of reference $r(f,t)$ is adjusted in each frequency bin to satisfy $\langle r(f,t)^2 \rangle_t = 1$.

#### 1) TV Gaussian Model

The TV Gaussian model is a type of TV distribution model that contains different variances in each frequency bin and frame. It is widely used in BSS problems [15][16]. To represent the dependence between the estimated target source $y_1(f,t)$ and reference $r(f,t)$, we interpret $r(f,t)$ as a value related to the variance. To control the influence of the reference, we append $\beta$ as the *reference exponent* to the original TV Gaussian model:

$$p(r(f,t), y_1(f,t)) \propto \frac{1}{r(f,t)^{\beta/2}} \exp\left(-\frac{|y_1(f,t)|^2}{r(f,t)^\beta}\right). \quad (17)$$

The strict TV Gaussian model corresponds to the case $\beta = 2$.

Next, we derive the update rule for the TV Gaussian model. Assigning (17) to (15) leads to the following minimization problem under Constraint (16):

$$\boldsymbol{w}_1(f) = \arg\min_{\boldsymbol{w}_1(f)} \left\{ \langle \frac{|y_1(f,t)|^2}{r(f,t)^\beta} \rangle_t \right\}. \quad (18)$$

This problem has the following closed-form solution using the eigenvalue decomposition [19]:

$$\boldsymbol{w}_1(f) = \text{EIG}_{\min}\left( \langle \frac{\boldsymbol{u}(f,t)\boldsymbol{u}(f,t)^{\mathrm{H}}}{r(f,t)^\beta} \rangle_t \right), \quad (19)$$

where $\text{EIG}_{\min}(\cdot)$ denotes the eigenvector corresponding to the minimum eigenvalue of the given matrix. To prevent division by zero, we use the following formulas instead of (19):

$$r'(f,t) = \max(r(f,t)^\beta, \ \varepsilon), \quad (20)$$

$$\boldsymbol{w}_1(f) = \text{EIG}_{\min}\left( \langle \frac{\boldsymbol{u}(f,t)\boldsymbol{u}(f,t)^{\mathrm{H}}}{r'(f,t)} \rangle_t \right), \quad (21)$$

where $\max(\cdot)$ denotes the selection of the maximum argument, and $\varepsilon$ is a small positive value referred to as a *clipping threshold*.

#### 2) BS Laplacian Model

The BS models, including the BS Laplacian model, are a two-variable version of the multivariate spherical (MS) distribution [20][21][22]. Considering that the MS models have a characteristic that the components in the square root gain mutual dependence, we use the following BS Laplacian model to make $|y_1(f,t)|$ and $r(f,t)$ mutually dependent:

$$p(r(f,t), y_1(f,t)) \\ \propto \exp\left(-\sqrt{\alpha r(f,t)^2 + |y_1(f,t)|^2}\right), \quad (22)$$

where $\alpha$ is a nonnegative value and referred to as the *reference weight*. This weight can control the influence of the reference. A larger value of $\alpha$ makes $|y_1(f,t)|$ more negligible. In contrast, the case of $\alpha = 0$ is equivalent to (23).

$$\mathrm{p}(y_1(f,t)) \propto \exp(-|y_1(f,t)|). \tag{23}$$

Given that (23) represents the source model in the conventional frequency-domain ICA using the univariate Laplacian distribution [23], the case of $\alpha = 0$ corresponds to the extraction using only the independence, in other words, without using the dependence.

The derivation of the update rules for the BS Laplacian model are presented below. Assigning (22) to (15) leads to the following minimization problem under Constraint (16):

$$\boldsymbol{w}_1(f) = \arg\min_{\boldsymbol{w}_1(f)} \left\{ \langle \sqrt{\alpha r(f,t)^2 + |y_1(f,t)|^2} \rangle_t \right\}. \tag{24}$$

This problem can be solved stably and efficiently using the auxiliary function algorithm [24][25]. To apply this algorithm, we consider the following inequality, which contains a positive value $b(f,t)$ referred to as the *auxiliary variable*:

$$\begin{aligned}
&\sqrt{\alpha r(f,t)^2 + |y_1(f,t)|^2} \\
&\leq \frac{\alpha r(f,t)^2 + |y_1(f,t)|^2}{2b(f,t)} + \frac{b(f,t)}{2}.
\end{aligned} \tag{25}$$

This inequality indicates that the arithmetic mean of the two terms on the right-hand side is greater than or equal to their geometric mean, which is the left-hand-side term.

We derive a sequence of iterative rules that consist of (7), (26), (27), and (28).

$$b(f,t) \leftarrow \sqrt{\alpha r(f,t)^2 + |y_1(f,t)|^2}, \tag{26}$$

$$b'(f,t) \leftarrow \max(b(f,t), \varepsilon), \tag{27}$$

$$\boldsymbol{w}_1(f) \leftarrow \mathrm{EIG}_{\min}\left( \langle \frac{\boldsymbol{u}(f,t)\boldsymbol{u}(f,t)^{\mathrm{H}}}{b'(f,t)} \rangle_t \right). \tag{28}$$

To emphasize that these formulas are the assignment in the iterative process, we use the left arrow "$\leftarrow$" instead of "$=$," as needed.

In the first iteration, both $\boldsymbol{w}_1(f)$ and $y_1(f,t)$ are unknown. Therefore, we derive the initial rule as follows. Regarding $r(f,t)$ as an estimate of $|y_1(f,t)|$ in (26) leads to the rule $b(f,t) \leftarrow r(f,t)\sqrt{\alpha+1}$. Given that the multiplication of $\sqrt{\alpha+1}$ does not influence the extraction filter, this rule is equivalent to the TV Gaussian model with $\beta = 1$. Therefore, we use (20) and (21) with $\beta = 1$ in the first iteration to obtain identical results to those obtained by the TV Gaussian model.

The update rules for the BS Laplacian model also help us to understand the case of $\alpha \to \infty$. Considering that $|y_1(f,t)|$ is negligible in (26), this formula is approximated as

$b(f,t) \leftarrow r(f,t)\sqrt{\alpha}$. Similar to the first iteration, this corresponds to the rules for the TV Gaussian model with $\beta = 1$.

### 3) TV t Model

In this study, we examine the TV t model, which is another variation of the time-frequency-varying distribution, and is used to solve BSS problems [16][26]. The TV t model is expressed as follows:

$$\mathrm{p}(r(f,t), y_1(f,t)) \propto \frac{1}{r(f,t)^2}\left(1 + \frac{2}{\nu}\frac{|y_1(f,t)|^2}{r(f,t)^2}\right)^{-\frac{2+\nu}{2}}, \tag{29}$$

where $\nu$ is the degree of freedom required to control the shape of the distribution; the case $\nu = 1$ corresponds to the TV complex Cauchy distribution, and the case $\nu \to \infty$ is equivalent to the TV Gaussian model (17) with $\beta = 2$.

The derivation of the update rules for the TV t model is presented below. Assigning (29) to (15) leads to the following minimization problem under Constraint (16):

$$\boldsymbol{w}_1(f) = \arg\min_{\boldsymbol{w}_1(f)} \left\{ \langle \log\left(1 + \frac{2}{\nu}\frac{|y_1(f,t)|^2}{r(f,t)^2}\right) \rangle_t \right\}. \tag{30}$$

To apply the auxiliary function algorithm to (30), we consider the following inequality containing the auxiliary variable $b(f,t)$ [26]:

$$\begin{aligned}
&\log\left(1 + \frac{2}{\nu}\frac{|y_1(f,t)|^2}{r(f,t)^2}\right) \\
&\leq \frac{1}{b(f,t)}\left(1 + \frac{2}{\nu}\frac{|y_1(f,t)|^2}{r(f,t)^2}\right) - 1 + \log(b(f,t)).
\end{aligned} \tag{31}$$

Similar to [16], we use $\xi(f,t) = \frac{\nu}{\nu+2}b(f,t)r(f,t)^2$ instead of $b(f,t)$ and derive a sequence of iterative rules consisting of (7), (32), (33), and (34).

$$\xi(f,t) \leftarrow \frac{\nu}{\nu+2}r(f,t)^2 + \frac{2}{\nu+2}|y_1(f,t)|^2, \tag{32}$$

$$\xi'(f,t) \leftarrow \max(\xi(f,t), \varepsilon), \tag{33}$$

$$\boldsymbol{w}_1(f) \leftarrow \mathrm{EIG}_{\min}\left( \langle \frac{\boldsymbol{u}(f,t)\boldsymbol{u}(f,t)^{\mathrm{H}}}{\xi'(f,t)} \rangle_t \right). \tag{34}$$

In the first iteration, we regard $r(f,t)$ as an estimate of $|y_1(f,t)|$. This leads to the rule $\xi(f,t) \leftarrow r(f,t)^2$, which is equivalent to the TV Gaussian model with $\beta = 2$.

From a comparison between (32) and (26), the equations are found to be similar. The main difference is in whether the square root is applied or not. Therefore, we interpret $\nu$ as a value like the reference weight in the BS Laplacian model, and thus consider the case of $0 < \nu < 1$. The case of $\nu \to 0$ is equivalent to the following complex Cauchy distribution with $\rho \to 0$:

| **Algorithm 1**: Iterative algorithm of SIBF with the boost start |
|---|
| **Input**: |
| $\quad$ $\boldsymbol{U}_1, \dots, \boldsymbol{U}_N$: Uncorrelated observation spectrograms of $N$ microphones |
| $\quad$ $\boldsymbol{R}$: Magnitude spectrogram as reference |
| **Output**: Spectrogram of the estimated target source |
| 1:   **function** SIBF_with_boost_start($\boldsymbol{U}_1, \dots, \boldsymbol{U}_N, \boldsymbol{R}$) |
| 2:     **for** $l = 1$ to $L_{\text{filter}}$ **do** |
| 3:       **for** all frequency bin $f$ **do** |
| 4:         **if** $l = 1$ **then** // Use boost start in 1st iteration |
| 5:           Calculate $\boldsymbol{w}_1(f)$ using (20) and (21) with $\beta_{\text{best}}$. |
| 6:         **else** // After 2nd iteration |
| 7:           Update $y_1(f, t)$ using (7) for all $t$. |
| 8:           Update $\boldsymbol{w}_1(f)$ using (26) to (28) for BS Laplacian model, or (32) to (34) for TV t model. |
| 9:         **end if** |
| 10:       **end for** |
| 11:     **end for** |
| 12:     **return** $Y_1$ |
| 13:   **end function** |

| **Algorithm 2**: Algorithm of iterative casting |
|---|
| **Input**: |
| $\quad$ $\boldsymbol{X}_1, \dots, \boldsymbol{X}_N$: Observation spectrograms of $N$ microphones |
| $\quad$ $m$: Microphone index used in the scaling |
| **Output**: Spectrogram of the SIBF output in $L_{\text{cast}}$-th casting |
| 1:   **function** SIBF_with_iterative_casting($\boldsymbol{X}_1, \dots, \boldsymbol{X}_N, m$) |
| 2:     // Initialization |
| 3:     **for** all frequency bin $f$ **do** |
| 4:       Calculate $\boldsymbol{u}(f, t)$ using (1) for all $t$. |
| 5:     **end for** |
| 6:     $Y_1 \leftarrow X_m$ |
| 7:     // Loop for casting |
| 8:     **for** $l = 1$ to $L_{\text{cast}}$ **do** |
| 9:       // Reference generation and normalization |
| 10:       $\boldsymbol{R} \leftarrow \text{DNN}(|Y_1|)$ |
| 11:       **for** all frequency bin $f$ **do** |
| 12:         Adjust scale of the reference so that $\langle r(f,t)^2 \rangle_t = 1$. |
| 13:       **end if** |
| 14:       // Filter estimation |
| 15:       Update $\boldsymbol{w}_1(f)$ for all $f$ using (20) and (21) for TV Gaussian model, or using Algorithm 1 for BS Laplacian and TV t models. |
| 16:       // Scaling |
| 17:       **for** all frequency bins $f$ **do** |
| 18:         Update $y_1(f, t)$ using (7) for all $t$. (only for TV Gaussian model) |
| 19:         Calculate $\gamma_1(f)$ using (37). |
| 20:         Update $y_1(f, t) \leftarrow \gamma_1(f) y_1(f, t)$ for all $t$. |
| 21:       **end for** |
| 22:     **end for** |
| 23:     **return** $Y_1$ |
| 24:   **end function** |

$$\mathrm{p}(y_1(f, t)) \propto \frac{\rho}{(|y_1(f, t)|^2 + \rho^2)^{3/2}}, \qquad (35)$$

where $\rho$ is the scale parameter that specifies the range of the distribution.

### D. Scaling in Postprocess

After filter estimation, we estimate the appropriate scales and phases of the estimated target source. This process is the same as that in the conventional frequency-domain ICA, such as the minimal distortion principle [27][28]. It is conducted by minimizing the following approximate error:

$$\langle \left| x_m(f, t) - \sum_k \gamma_k(f) y_k(f, t) \right|^2 \rangle_t, \qquad (36)$$

where $x_m(f, t)$ and $\gamma_k(f)$ denote a scaling reference, which is the observation of the $m$th microphone and a scaling factor for $y_k(f, t)$, respectively. From the mutual independence of the estimated sources and $\langle |y_1(f, t)|^2 \rangle_t = 1$, $\gamma_1(f)$ is calculated as follows:

$$\gamma_1(f) = \langle x_m(f, t) \overline{y_1(f, t)} \rangle_t, \qquad (37)$$

where $\overline{y_1(f, t)}$ denotes the conjugate of $y_1(f, t)$.

The estimated target source after the scaling is expressed as $\gamma_1(f) y_1(f, t)$, which is referred to the *SIBF output*.

There is another advantage in combining the SIBF and scaling, namely the capability of estimating the appropriate phases from the given magnitude spectrogram $\boldsymbol{R}$. This indi-cates that even if the reference is highly accurate as a magnitude spectrogram, the SIBF is useful for estimating the corresponding phases.

## III. METHODS FOR PERFORMANCE IMPROVEMENT

In this section, we propose two methods for improving the extraction performance of the SIBF.
1. Acceleration of the convergence, which is referred to as *boost start*.
2. Feedback of the SIBF output to the DNN, which is referred to as *iterative casting*.

A process such as iterative casting has been used in both independent deeply learned matrix analysis (IDLMA) [16] for the reference-based source separation and iterative mask estimation [12][13] for the mask-based BF. However, this study is unique in the flowing aspects:

- This study applies iterative casting to the target source extraction problem, unlike IDLMA, and presents a precise discussion on the relationship between the accuracies of the reference and SIBF output in each count of casting.

- In this study, the DNN output after iterative casting can be directly used as an updated reference. In [12] and [13], however, estimating an updated mask was more complicated; it required combining the DNN output and other masks estimated by other methods such as those employing the complex Gaussian mixture model and ASR-based voice activity detector.

### A. Boost Start

As mentioned in Section II.C, the initial rule in the first iteration should be considered, given that both $w_1(f)$ and $y_1(f, t)$ are unknown. In that section and in our previous study [8], $r(f, t)$ was regarded as an approximation of $|y_1(f, t)|$. This interpretation leads to initial rules that are equivalent to the TV Gaussian model with $\beta = 1$ (for the BS Laplacian model) and $\beta = 2$ (for the TV t model). We refer to the use of these rules as *model-specific start*.

In this study, we propose an alternative method called *boost start*, to generate $y_1(f, t)$ using a more effective method. For the BS Laplacian and TV t models, we can achieve boost start using $\beta_{best}$, which provides the optimal accuracy for the TV Gaussian model. Algorithm 1 presents the pseudocode for the SIBF using the boost start, and $L_{filter}$ denotes the maximum iteration counts for the filter estimation. It should be noted that calculating $y_1(f, r)$ updates $Y_1$ because $y_1(f, r)$ is an element of $Y_1$. We assume that the reference $R$ is appropriately scaled, and the estimated target source is scaled after calling this function.

### B. Iterative casting

As discussed in our previous study [8], the use of a more accurate reference can lead to a more accurate extraction of the target source. At minimum, there are two options for this: refining the DNN for reference generation and casting the SIBF output to the DNN. This study examines the latter option, based on the assumption that if the current DNN input is more accurate (closer to the target source) than the previous input, the corresponding output can be more accurate than the previous output.

We define iterative casting as follows:

**Initial casting**: Casting part of the observation spectrograms into the DNN, as mentioned in Section II.

**n-th casting** ($n > 1$): Casting the previous SIBF output, which was generated in the previous casting, into the DNN.

To distinguish between the iterative casting and iterative update rules for the filter estimation, we use the term *iteration* only for the filter estimation. We refer to the reference and SIBF output in the initial casting as the *initial reference* and *initial SIBF output*, respectively, and those in the *n*-th casting as the *n*-th reference and *n*-th SIBF output, respectively.

TABLE II
Experimental setups described in each subsection ("DNN" and "Oracle" indicate a reference generated by DNN and the ideal one, respectively. Dataset is explained in Section IV.B.)

| Sub-section | Methods | Reference type | Dataset | Scores |
|---|---|---|---|---|
| IV.C | Model-specific start | DNN & Oracle | Develop | PESQ |
| IV.D | Boost start | DNN | Develop | PESQ |
| IV.E | Boost start & Iterative casting | DNN | Develop | PESQ, SDR |
| IV.F | Boost start & Iterative casting | DNN & Oracle | Test | PESQ, SDR, STOI |

Algorithm 2 presents the pseudocode for iterative casting, where $L_{cast}$ denotes the maximum count of casting, and DNN($\cdot$) is the function to obtain the DNN output by casting the given data into the DNN. Algorithm 2 uses the boost start in the filter estimation for the BS Laplacian and TV t models; it also includes the decorrelation (lines 3–5), reference generation (line 10), reference normalization (lines 11–13), and scaling (lines 17–21).

## IV. Experiments

To verify the effectiveness of the SIBF and its improvement, we conducted a series of experiments using the CHiME3 dataset [29]. This dataset contains sound data recorded in four noisy environments using six microphones attached to a tablet device and clean speech recorded in a recording booth. The use of this dataset means that the SIBF is applied to a denoising task, in which speech is extracted from a mixture of speech and background noise.

The experiments consisted of the following four steps:
1. Tuning the hyperparameters for each source model
2. Evaluating the effects of the boost start for the BS Laplacian and TV t models
3. Evaluating iterative casting for each source model
4. Comparing the results obtained using the CHiME3 simulated test set with those of other methods

In this section, we first explain the DNN for reference generation as well as the dataset used for the development and evaluations. Then, each experimental step listed in Table II is detailed in the following subsections.

### A. DNN Configuration

To prepare the DNN for reference generation, we modified the configuration that trains the bidirectional long short-term memory (BLSTM) based mask estimator for the mask-based Max SNR BF [9], which is included in the CHiME4 speech enhancement baseline [1], to output the magnitude spectrogram. The network configuration is shown in Fig. 3, which is the same as that in our previous study [8]. This DNN was trained to extract a single speech source from a mixture of a single speech source and one or more nonspeech sources. The differences between the configuration in this study and that in [9] were limited to the following aspects.

1. The supervisory data consisted of magnitude spectrograms of clean speech instead of ideal binary masks.
2. The mean squared error (MSE) was used instead of the cross-entropy loss as the loss function in the training stage.
3. The training was performed in 20 epochs.

It should be noted that the modification of the DNN configuration was minimal because the objective of the experiments was to evaluate the effectiveness of the SIBF.

### B. Dataset and Common Setups

To prepare the input data as a development set with various signal-to-noise ratios (SNRs), we artificially mixed the clean speech with background (BG) noise. The clean speech consisted of 410 utterances from four speakers and were recorded in a booth using six microphones attached on a tablet device. This dataset is labeled as *dt05_bth* in the CHiME3 webpage [1]. Background noises were recorded with the same device in four noisy environments such as a bus, café, pedestrian area, and street junction. In the CHiME3 webpage, these datasets are labeled as BUS, CAF, PED, and STR, respectively. During the mixing, we applied four multipliers (0.25, 0.5, 1.0, and 2.0) to the noises to represent the four scenarios, as shown in Table III.

In Section IV.F, we used another dataset called *CHiME3 simulated test set* for comparing the SIBF with other methods. This consists of a combination of 330 utterances from four speakers and four background noises.

In the preprocessing, we converted the waveforms into spectrograms using the short-time Fourier transform (STFT) with 1024 points and 256 shifts. Thus, the total number of frequency bins $F = 513$. In the postprocess, we employed (37) with $m = 5$, which was the index of the microphone closest to the speaker. We employed $\varepsilon = 10^{-7}$ as the clipping threshold in (20), (27), and (33). Before applying the inverse STFT (ISTFT), we replaced the frequency bin data under 62.5 Hz and over 7812.5 Hz with zeros, because the training and evaluation system used in [9] includes processes with the same effects. This process was also applied to the ISTFT for the evaluation of the reference, whereas this was not applied in the evaluation of the observation.

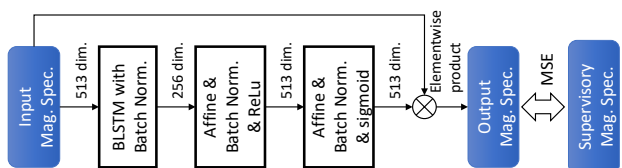

Fig. 3: Network configuration for the DNN that outputs a magnitude spectrogram (Mag. Spec.) as the reference. The numbers indicate the input and output dimensions.

TABLE III
FOUR SCENARIOS USED AS THE DEVELOPMENT SET

| Scenario name | Relative noise level | Multiplier to BG | PESQ | SNR [dB] |
|---|---|---|---|---|
| BG × 0.25 | Least noisy | 0.25 | 2.93 | 14.05 |
| BG × 0.5 | Less noisy | 0.5 | 2.51 | 8.03 |
| BG × 1.0 | Noisier | 1.0 | 2.10 | 2.03 |
| BG × 2.0 | Noisiest | 2.0 | 1.72 | -3.93 |

### C. Hyperparameter Tuning

To determine the optimal hyperparameters for each model, we conducted a series of experiments using the perceptual evaluation of speech quality (PESQ) [30] as the performance score, considering that the PESQ is extensively employed in studies that use the CHiME3 dataset.

Fig. 4 illustrates the evaluation system used in the experiments. This system can switch the type of reference that is used. One is a DNN-based reference, which is generated by the DNN from the observation of Microphone #5. The other is the Oracle reference, which is the magnitude spectrogram of clean speech before mixing. The SIBF using the DNN-based reference is referred to as the *NN SIBF*, and that using the Oracle reference is referred to as the *Oracle SIBF*. The objective of evaluating the Oracle SIBF was to estimate the potential performance of the SIBF when an ideal reference was provided.

In the experiment using the NN SIBF, we evaluated the reference as a waveform, in addition to the SIBF output, by combining the DNN output and observation phases and applying the ISTFT to it.

The results for each source model are discussed below. Although we conducted the experiments using the TV Gaussian and BS Laplacian models in our previous study [8], in this study, we re-examined these models with wider ranges of the hyperparameters and obtained the results for all scenarios.

#### 1) Tuning for TV Gaussian Model

The TV Gaussian model (17) contains the reference exponent $\beta$ as a hyperparameter. We evaluated the SIBF using this model within the range of $1/8 \le \beta \le 32$.

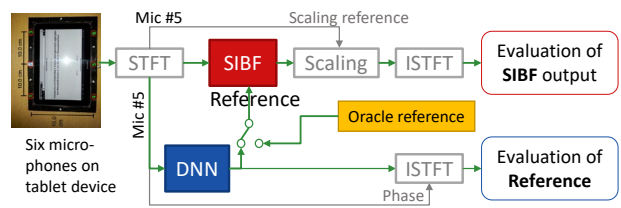

Fig. 4: Evaluation system for SIBF (without iterative casting) using CHiME3 dataset

Fig. 5 presents a subset of the PESQ scores of the NN SIBF: BG × 0.25 (left) and BG × 2.0 (right) scenarios. The score of the reference is also plotted as a dashed line for comparison. The results for the BG × 0.5 and BG × 1.0 scenarios are not plotted, considering that these demonstrate almost the same trends as those for BG × 0.25 and BG × 2.0 scenarios, respectively. We found that the case of $\beta = 2$ yielded the lowest score among all the scenarios, although only this case strictly corresponds to the TV Gaussian distribution. The PESQ scores improved as $\beta$ moved away from 2 in both directions. However, under the BG × 0.25 and BG × 0.5 scenarios, the cases of $\beta = 1/2$ and $\beta = 8$ demonstrated a peak.

Fig. 6 presents the scores of the Oracle SIBF. The trends in both figures are highly similar; however, the scores in Fig. 6 are higher than those in Fig. 5 except for the cases of $\beta \leq 1/2$.

From these results, we selected $\beta = 8$ as the optimal hyperparameter, as it yielded almost the highest performance in all the scenarios. This hyperparameter was also employed as $\beta_{best}$ for the boost start, as mentioned in Section III.A.

#### 2) Tuning for BS Laplacian Model

The BS Laplacian model (22) contains two hyperparameters, namely the reference weight $\alpha$ and the maximum iteration count. We evaluated the relationship between the PESQ scores and iteration counts (1–20) for four reference weights: $\alpha = 0.01, 1, 100,$ and $10^4$.

Fig. 7 presents a subset of the results of the NN SIBF and references: BG × 0.25 (left) and BG × 2.0 (right) scenarios. The results for the BG × 0.5 and BG × 1.0 scenarios are not plotted, considering that these demonstrate almost the same trends as those for BG × 0.25 and BG × 2.0 scenarios, respectively. As mentioned in Section II.C.2, the first iteration is substantively equivalent to the TV Gaussian model with $\beta = 1$. We found that different $\alpha$ values resulted in different trends. For example, the case of $\alpha = 0.01$ showed the most significant improvement under the BG × 2.0 scenario as the iteration count increased and showed significant degradation under the BG × 0.25 scenario. The case of $\alpha = 10^4$ improved slightly in all the scenarios.

Fig. 8 presents the results for the Oracle SIBF. We found that the cases of both $\alpha = 100$ and $\alpha = 10^4$ improved only slightly in all the scenarios, whereas the case of $\alpha = 0.01$ degraded.

From these results, we selected $\alpha = 100$ as the optimal hyperparameter, considering that it exhibited a stable increasing trend in all the scenarios.

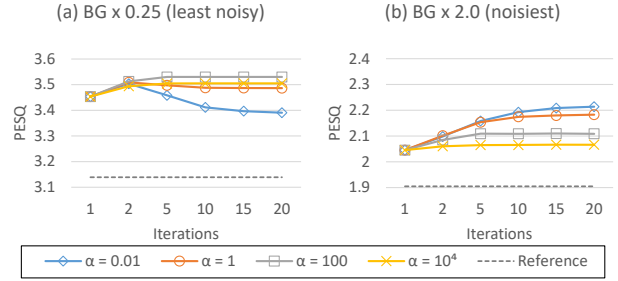

Fig. 7: PESQ scores of the NN SIBF using the BS Laplacian model

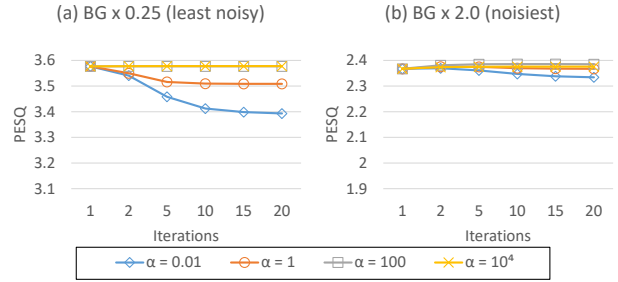

Fig. 8: PESQ scores of the Oracle SIBF using the BS Laplacian model

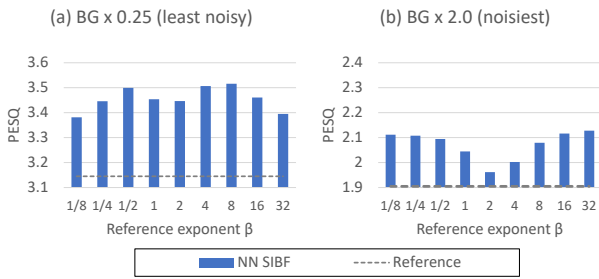

Fig. 5: PESQ scores of the NN SIBF using the TV Gaussian model

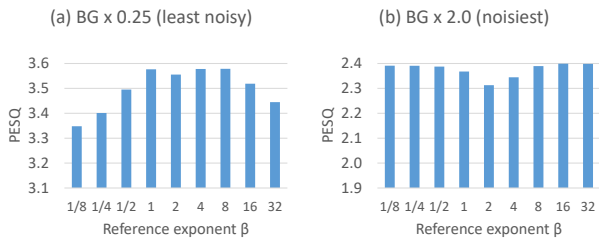

Fig. 6: PESQ scores of the Oracle SIBF using the TV Gaussian model

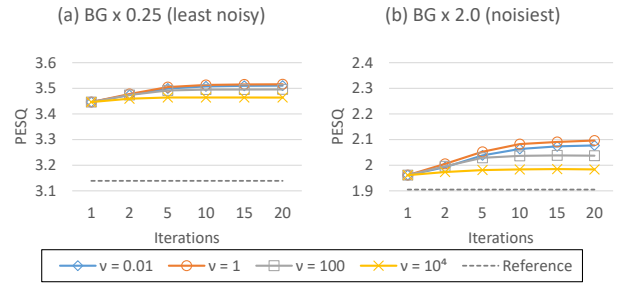

Fig. 9: PESQ scores of the NN SIBF using the TV t model

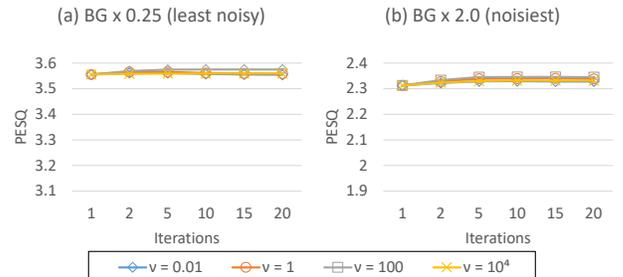

Fig. 10: PESQ scores of the Oracle SIBF using the TV t model

### 3) Tuning for TV t Model

The TV t model (29) contains two hyperparameters, namely, the degree of freedom $\nu$ and the maximum iteration counts. Similar to the experiments using the BS Laplacian model, we evaluated the relationship between the PESQ score and iteration count for four degrees of freedom: $\nu = 0.01, 1, 100,$ and $10^4$.

Fig. 9 presents a subset of the results of the NN SIBF and references: BG × 0.25 (left) and BG × 2.0 (right) scenarios. The results for the other scenarios are not plotted, considering that these almost the same trend as Fig. 9. As mentioned in Section II.C.3, the first iteration was equivalent to the TV Gaussian model with $\beta = 2$. Unlike that in the BS Laplacian model, we found that the case of $\nu = 1$ commonly resulted in the most significant improvement in all the scenarios.

Fig. 10 presents the results of the Oracle SIBF. All the scenarios exhibited the same trend in that the PESQ scores improved only slightly regardless of the degree of freedom $\nu$.

From these results, we selected $\nu = 1$ as the optimal hyperparameter.

### 4) Discussion on Hyperparameter Tuning

Both Fig. 5 and Fig. 6 suggest that for the TV Gaussian model, the extraction performance is sensitive to the reference exponent $\beta$. The case of $\beta = 2$ may output poor results, although this value is extensively used in studies employing this model. Therefore, further clarification of the sensitivity characteristic is necessary. Section VI.B presents a discussion on the sensitivity using the unified formulation of the SIBF and mask-based Max SNR BF.

Both Fig. 7 and Fig. 8 suggest that for the BS Laplacian model, the optimal value of the reference weight $\alpha$ is determined by the accuracy of the reference. For the Oracle SIBF, the reference is ideally accurate regardless of the noisiness of the observation. Thus, a larger $\alpha$ results in improved scores in all the scenarios. However, for the NN SIBF, the accuracy of the reference is considered significantly dependent on the noisiness of the observation. For example, in the BG × 2.0 scenario, the reference was less accurate than that in the other scenarios because the observation was noisier. Thus, the case of $\alpha = 0.01$ resulted in the most significant improvement of the scores. In contrast, in the BG × 0.25 scenario, the reference was more accurate than that in the other scenarios but less accurate than that of Oracle SIBF. Thus, the case of $\alpha = 100$, which is larger than 0.01 but less than $10^4$, showed the most significant improvement in the scores.

Remarkably, we found that the SIBF using the BS Laplacian model with $\alpha = 0.01$ converges to an almost identical value in the same scenario for both DNN-based and Oracle references. This trend typically appears in the BG × 0.25 scenario, as shown in both Fig. 7 and Fig. 8, and can be explained as follows. The weight $\alpha = 0.01$ is so close to zero that the constraint by the reference rarely works except in the first iteration. Therefore, as the iteration counts increase, the score converges to a value that can be obtained without using the reference.

For the TV t model, the case of $\nu = 1$ commonly yielded optimal scores in all the scenarios, as shown in Fig. 9, which was different from the BS Laplacian model. This suggests that the optimal value of $\nu$, which is the degree of freedom, is not linked to the accuracy of the reference, and represents the distribution of the target source. This may be because the clean speech data used as the target source in the experiments were in accordance with a distribution similar to the TV Cauchy distribution, which corresponds to the case of $\nu = 1$ in the TV t model.

As mentioned in Section II.C.3, the update rules for both the BS Laplacian and TV t models, e.g., (26) and (32), have similar expressions. However, the experimental results indicated that the behaviors of both models were significantly different, although the main difference between both rules is in whether the square root is applied or not. Therefore, by clarifying the cause of this difference, more effective update rules can be realized.

### D. Boost Start vs. Model-specific Start

We evaluated the effectiveness of the boost start by comparing it with the model-specific start evaluated in the previous subsection. This was achieved using $\beta_{\text{best}} = 8$ in the first iteration for both the BS Laplacian and TV t models. In this subsection, the results of the NN SIBF are plotted just for the BG × 0.25 and BG × 2.0 scenarios, considering that the result for all the scenarios demonstrate the same trends. The results of the Oracle SIBF are not plotted, as they are highly similar to those for the model-specific start shown in Fig. 8 and Fig. 10.

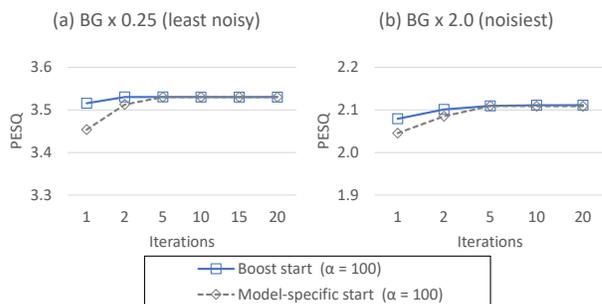

Fig. 11: Boost start (solid lines) vs. model-specific start (dotted lines) for the BS Laplacian model

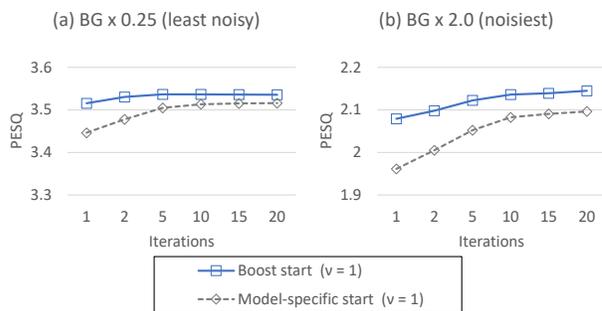

Fig. 12: Boost start (solid lines) vs. model-specific start (dotted lines) for the TV t model

TABLE IV
PESQ SCORES OF ALL THE THREE SOURCE MODELS (OPTIMAL SCORE FOR EACH SCENARIO IS IN BOLD TYPE)

| Source model | Hyper-parameters | Boost start | BG ×0.25 | BG ×0.5 | BG ×1.0 | BG ×2.0 |
|---|---|---|---|---|---|---|
| **BS Laplacian** | $\alpha = 100$, 20 iterations | | 3.53 | 3.13 | 2.66 | 2.11 |
| | | ✓ | 3.53 | 3.13 | 2.66 | 2.11 |
| **TV t** | $\nu = 1$, 20 iterations. | | 3.52 | 3.12 | 2.64 | 2.10 |
| | | ✓ | **3.54** | **3.15** | **2.68** | **2.14** |
| **TV Gaussian** | $\beta = 1$ | (n/a) | 3.45 | 3.04 | 2.58 | 2.04 |
| | $\beta = 2$ | | 3.45 | 3.00 | 2.49 | 1.96 |
| | $\beta = 8$ | | 3.52 | 3.12 | 2.63 | 2.08 |

### 1) The BS Laplacian Model with Boost Start

Fig. 11 presents the results of the NN SIBF using the BS Laplacian model with $\alpha = 100$. In the first iteration for each scenario, the scores with the boost start and model-specific start are substantively identical to those using the TV Gaussian model with $\beta = 8$ and $\beta = 1$, respectively.

We found that the effectiveness of the boost start was limited until the fifth iteration. However, the boost start was advantageous in terms of decreasing the maximum iteration count, given that the scores in the second iteration were almost identical to those in the 20th iteration. We also confirmed that the boost start did not change the convergence value.

### 2) The TV t Model with Boost Start

Fig. 12 presents the results of the NN SIBF using the TV t model with $\nu = 1$. In the first iteration for each scenario, the scores with the boost start and model-specific start are substantively identical to those using the TV Gaussian model with $\beta = 8$ and $\beta = 2$, respectively.

We found that even in the 20th iteration, the scores for the model-specific start did not converge; thus, the effectiveness of the boost start was maintained. Therefore, the boost start was advantageous for the TV t model in decreasing the maximum iteration count until convergence.

### 3) Discussion on Boost Start by Comparison with Three Source Models

Table IV presents the PESQ scores of all the three source models for all the scenarios. It should be noted that the TV Gaussian model with $\beta = 1$ and $\beta = 2$ is equivalent to the model-specific starts for the BS Laplacian and TV t models, respectively, and the case of $\beta = 8$ is equivalent to the boost start for both models. Thus, we can evaluate the relative changes after the 20th iteration.

When the model-specific start was used, the score of the TV t model in each scenario was observed to be lower than that of the BS Laplacian model. This is because of two aspects, namely lower score in the first iteration and slower convergence.

However, with the boost start, the scores of the TV t model were observed to be slightly higher than those of the BS Laplacian model. This is because the SIBF using the TV t model has a slightly higher convergence value in each scenario when compared with the BS Laplacian model, and the boost start allows for the SIBF score to reach this value in fewer iterations.

### E. Experiments on Iterative Casting

We then evaluated the effectiveness of iterative casting using the evaluation system shown in Fig. 13. This system can switch the signals casted to the DNN to generate the reference. One is the observation of Microphone #5, which is used only in the initial casting, and the other is the previous SIBF output. We also evaluated the reference as a waveform in each count of casting by combining it with the phases of the previous SIBF output and applying the ISTFT to the combined data. We used the signal-to-distortion ratio (SDR) [31] and PESQ as performance scores in the subsequent experiments. In this subsection, the results of iterative casting include those of the initial SIBF outputs, for convenience.

### 1) Results for Three Source Models

First, we examined the TV Gaussian model. Fig. 14 presents the results of iterative casting using the TV Gaussian model with $\beta = 8$. The vertical axes denote the PESQ for the top row and SDR for the bottom row; the horizontal axes commonly denote both the observation (obs.) and count of casting. This figure also includes two horizontal lines in each scenario, where the dashed line represents the score of the initial reference, and the dotted line represents that of the Oracle SIBF to indicate the maximum performance of the SIBF. In each scenario, the PESQ score of the initial SIBF output is substantively identical to that of the TV Gaussian model with $\beta = 8$ in Table IV.

As a common trend for all the scenarios, we found that the SIBF outputs after the second casting outperformed the initial reference and the observation, although the SDR scores in the BG ×0.25 scenario decreased to the initial reference. However, we found that the BG ×0.25 scenario yielded different behaviors from those of the other scenarios for the following two aspects:

1. In the BG ×0.25 scenario, the scores after the second casting were slightly degraded, whereas in the other

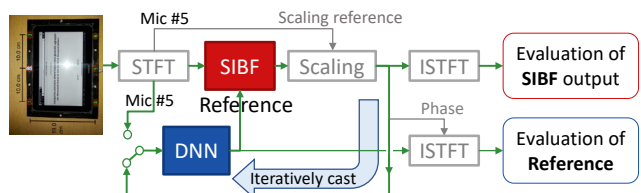

Fig. 13: Evaluation system for SIBF with iterative casting

scenarios, the scores improved as the casting count increased.

2. In the BG × 0.25 scenario, the PESQ scores were higher than those of the corresponding references, whereas the SDR scores were lower, except in the initial casting. However, in the other scenarios, the PESQ and SDR scores after the second casting were lower than those of the corresponding references.

Thereafter, we examined the iterative casting using the BS Laplacian and TV t models. The experimental procedures for both models were common. In each count of casting, the following steps were performed: (i) the reference was generated; (ii) the first iteration was performed with the boost start, which is equivalent to the TV Gaussian model with $\beta = 8$; (iii) the filter estimation was performed until 10th iteration, given that Fig. 11 and Fig. 12 suggest that the scores almost converged in the 10th iteration.

Only the results of the BS Laplacian model are shown in Fig. 15, because the results of the BS Laplacian and TV t models were highly similar. Unlike that in Fig. 14, the horizontal axis represents the count of casting and the iteration, and $n$-$m$ denotes the $m$th iteration in the $n$th casting, although this figure only presents the first and final (10th) iterations. The reference on $n$-1 is identical to that on $n$-10 because the reference is generated in each casting count. For each scenario, the PESQ scores in the initial casting, such as "1-1" and "1-10," are identical to those in Fig. 11. Similar to Fig.

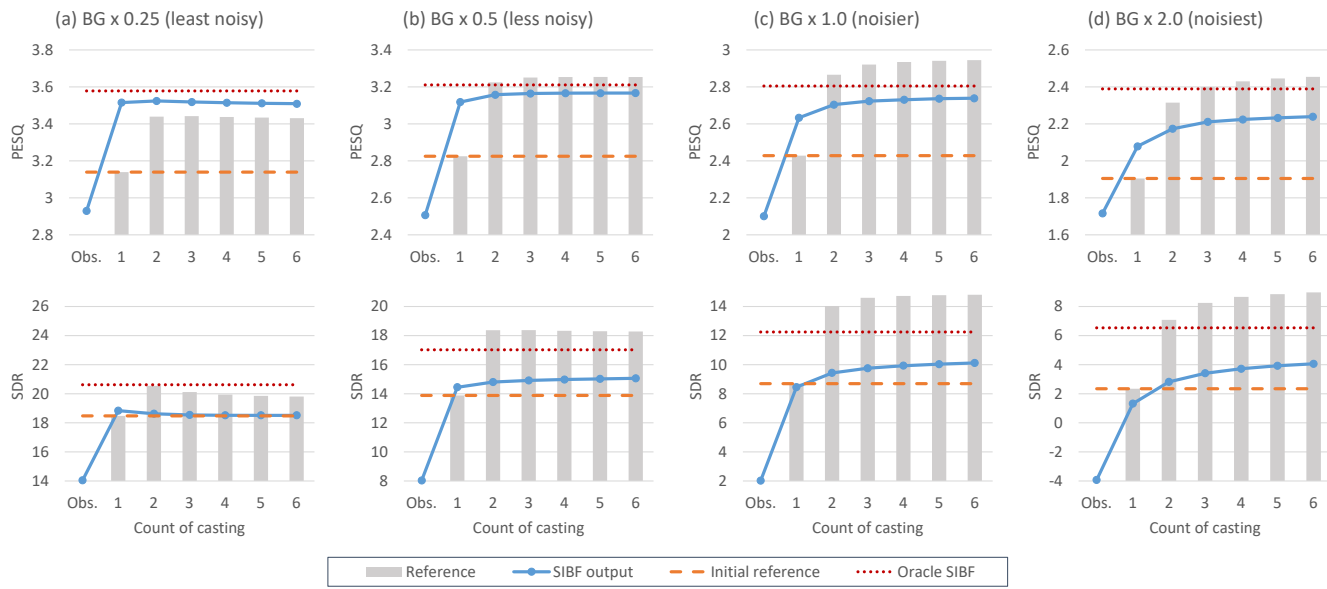

Fig. 14: Scores of iterative casting using the TV Gaussian model compared with the observation (obs.), references in each casting, and Oracle SIBF. The vertical axes denote the PESQ for the top row and SDR for the bottom row.

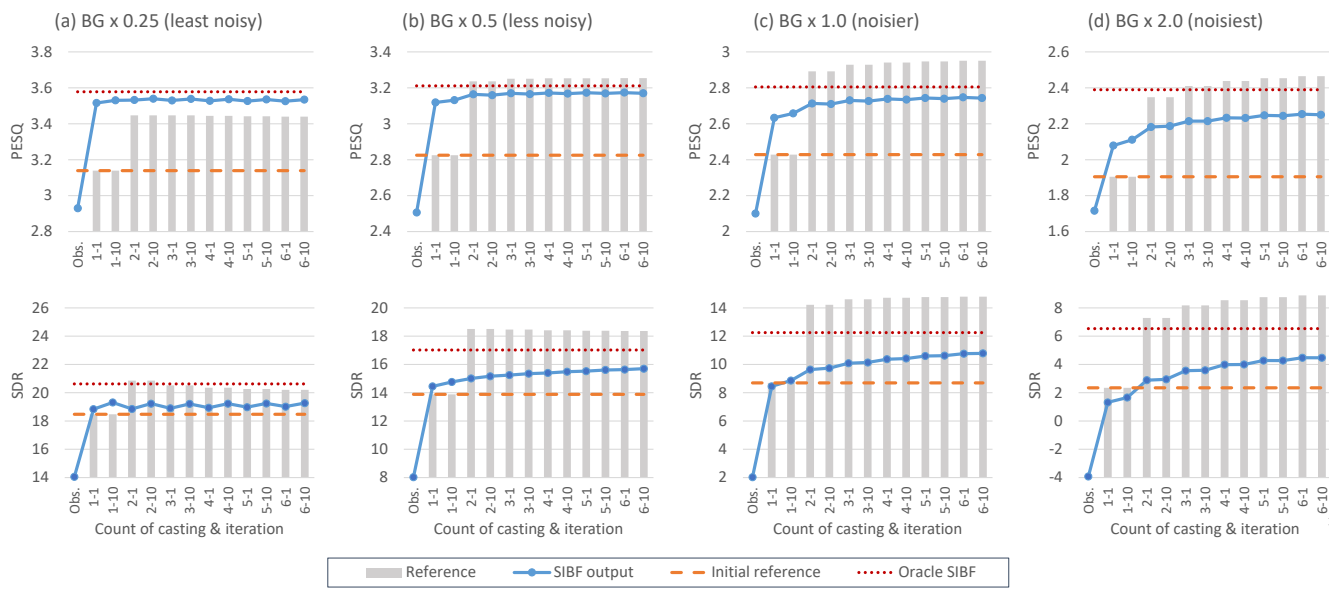

Fig. 15: Scores of iterative casting using the BS Laplacian model compared with the observation (obs.), references in each casting, and Oracle SIBF. The vertical axes denote the PESQ for the top row and SDR for the bottom row, and in the horizontal axis, $n$-$m$ represents the $m$-th iteration in the $n$-th casting.

TABLE V
PESQ AND SDR SCORES OF THE INITIAL REFERENCE, INITIAL AND SIXTH SIBF OUTPUTS, SIXTH REFERENCES, AND ORACLE SIBF (THE OPTIMAL SCORES IN EACH COLUMN ARE IN BOLD TYPE, ALTHOUGH THE ORACLE SIBF AND SIXTH REFERENCES WERE NOT CONSIDERED FOR THE COMPARISON)

| Signal type | Count of casting | Source model | PESQ | | | | SDR [dB] | | | |
|---|---|---|---|---|---|---|---|---|---|---|
| | | | BG ×0.25 | BG ×0.5 | BG ×1.0 | BG ×2.0 | BG ×0.25 | BG ×0.5 | BG ×1.0 | BG ×2.0 |
| Initial reference | 1 | (N/A) | 3.14 | 2.83 | 2.43 | 1.90 | 18.48 | 13.88 | 8.70 | 2.34 |
| SIBF outputs | 1 | TV Gaussian | 3.52 | 3.12 | 2.63 | 2.08 | 18.84 | 14.45 | 8.45 | 1.32 |
| | | BS Laplacian | 3.53 | 3.13 | 2.66 | 2.11 | 19.31 | 14.76 | 8.86 | 1.66 |
| | | TV t | **3.54** | 3.15 | 2.68 | 2.14 | **19.38** | 14.98 | 9.20 | 2.06 |
| | 6 | TV Gaussian | 3.51 | **3.17** | **2.74** | 2.24 | 18.52 | 15.06 | 10.12 | 4.06 |
| | | BS Laplacian | 3.53 | **3.17** | **2.74** | **2.25** | 19.27 | **15.70** | **10.78** | **4.47** |
| | | TV t | **3.54** | **3.17** | 2.73 | 2.23 | 19.28 | 15.38 | 10.22 | 3.88 |
| References in 6th casting | 6 | TV Gaussian | 3.43 | 3.25 | 2.94 | 2.45 | 19.81 | 18.28 | 14.81 | 8.98 |
| | | BS Laplacian | 3.44 | 3.25 | 2.95 | 2.47 | 20.20 | 18.36 | 14.80 | 8.88 |
| | | TV t | 3.45 | 3.26 | 2.95 | 2.45 | 20.45 | 18.44 | 14.78 | 8.87 |
| Oracle SIBF | (n/a) | TV Gaussian | 3.58 | 3.21 | 2.80 | 2.39 | 20.62 | 17.03 | 12.25 | 6.54 |

14, the scores of both the initial reference and the Oracle SIBF output using the TV Gaussian model were plotted.

We found that the results of the BS Laplacian model demonstrated the same trends as those of the TV Gaussian model. As a common trend for all the scenarios, the SIBF outperformed the initial reference in the PESQ and SDR after the second casting, and the BG × 0.25 scenario demonstrated different behaviors than the other scenarios for the aforementioned aspects.

In terms of the effect of iterations, the BS Laplacian model was unique. In the BG × 0.25 scenario, both the PESQ and SDR scores were improved by the iteration, although the scores were degraded by iterative casting. For example, "1-10" in Fig. 15 was higher than "1-1" due to the iteration, whereas "2-1" was lower than "1-10" due to iterative casting. In the other scenarios, however, the improvement by the iteration was almost limited to the initial casting. For example, "1-10" was higher than "1-1" in PESQ and SDR, whereas "2-10" was almost identical to "2-1," except for the SDR in the scenario of BG × 0.5.

Table V presents the results of the initial and sixth SIBF outputs over the three source models, in addition to the scores of the initial reference. For further discussion, the table includes the scores of the sixth reference and Oracle SIBF using the TV Gaussian model. The optimal scores in each column are presented in bold type, although the sixth references and Oracle SIBF were not considered for the comparison. As a common trend for all the source models, the scores of the sixth SIBF outputs were higher than those of the initial references, and the differences in sixth SIBF outputs among the models were dependent on the evaluation scores. For the PESQ, the differences were small, whereas for the SDR, the scores of the BS Laplacian model were significantly improved by iterative casting, except in the scenario of BG × 0.25.

As shown in Table V, the scores of the sixth reference were higher than those of the Oracle SIBF and the sixth SIBF outputs at several instances. These findings are significant for developing a better understanding of the behaviors of the SIBF, as discussed in Section V.

### 2) Discussion on Iterative Casting

A discussion on the causes of the findings is presented below:

1. The improvement of the scores by iterative casting.
2. The different behaviors demonstrated by the BG × 0.25 scenario at several instances in comparison with the other scenarios.
3. The inverse relationship between the changes in the SDR and those in the PESQ at several instances.

A further discussion on the behaviors of iterative casting is presented in Section V.

Regarding the first point, the score improvement can be explained based on the following assumptions:

• The accuracy of the DNN output monotonically increases with that of the DNN input, namely the observation or previous SIBF output.
• The accuracy of the SIBF output monotonically increases with that of the reference, namely the DNN output.

The degree of improvement per casting decreased as the accuracy of the SIBF output approached that of the Oracle SIBF output. This is because the SIBF output cannot exceed the Oracle SIBF output in principle. Moreover, for the BS Laplacian and TV t models, the degree of improvement by the iteration decreased as the accuracy of the reference increased. This can be explained by the fact that the results of the Oracle SIBF in Fig. 8 and Fig. 10 exhibited only a slight improvement despite the iteration.

The assumptions allow for the second point to be explained. In the BG × 0.25 scenario, the DNN outputs after the second casting were less accurate than the corresponding DNN inputs. This phenomenon can be interpreted as the elimination of the monotonicity assumption of the DNN by highly accurate inputs. We consider that it is due to the mismatch between the DNN input and training dataset; the SDRs of the SIBF outputs in this scenario were consistently greater

| Method | Count of casting | Source Model | PESQ | SDR [dB] | STOI [%] |
|--------|------------------|--------------|------|----------|----------|
| Initial reference | 1 | (n/a) | 2.61 | 13.61 | 91.50 |
| SIBF outputs (proposed) | 1 | TV Gaussian | 2.67 | 15.25 | 95.35 |
| | | BS Laplacian | 2.68 | 15.85 | 95.72 |
| | | TV t | 2.70 | 16.03 | 95.72 |
| | 6 | TV Gaussian | 2.71 | 16.76 | 96.07 |
| | | BS Laplacian | **2.72** | **17.29** | **96.18** |
| | | TV t | **2.72** | 16.81 | 96.04 |
| References in 6th casting | 6 | TV Gaussian | 2.99 | 19.61 | 96.32 |
| | | BS Laplacian | 3.00 | 19.62 | 96.36 |
| | | TV t | 3.00 | 19.60 | 96.30 |
| Oracle SIBF | (n/a) | TV Gaussian | 2.75 | 17.99 | 96.76 |
| Observation of Microphone #5 | | | 2.18 | 7.54 | 87.04 |
| Heymann et al. [1][9] (Mask-based Max SNR BF) | | | 2.46 | 2.92 | 87 |
| Erdogan et al. [11] (Mask-based MVDR BF) | | | 2.29 | 15.12 | |
| Tu et al. [12] (Mask-based MVDR BF with iterative mask estimation) | | | 2.71 | | 93.98 |

than 18 dB for all the source models, whereas the average SNR of the training dataset was approximately 5 dB [29].

A discussion on the third point is presented below. For example, in the initial casting for the BG × 2.0 scenario and iterative casting for the BG × 0.25 scenario, the SIBF outputs outperformed the corresponding reference in the PESQ, which was not the case for the SDR. In the sixth casting for the BG × 0.5 scenario, the PESQ score converged, whereas the SDR continuously increased. We consider that these phenomena can be caused by a combination of the following factors:

a) The SDR calculation is sensitive to residual interferences in the lower frequencies close to 0 Hz, whereas the PESQ calculation ignores such frequencies.

b) As a general characteristic of the BFs, the performance of removing the interferences in the lower frequency bins tends to be worse than that of removing those in the higher frequency bins because the phase differences between microphones become smaller at lower frequencies [32][33].

Ignoring the failure of the extraction in the lower frequency bins, we can regard that in the initial casting for the BG × 2.0 scenario and iterative casting for the BG × 0.25 scenario, the SIBF extracted the target source more accurately than the corresponding reference in most frequency bins. Given that iterative casting in the BG × 0.5 scenario mainly improved the SDR scores, we can consider that it mainly promoted the removal of the residual interferences in the lower bins.

### F. Comparison with Other Methods

To compare the performance of the SIBF with those of other methods, we conducted experiments using the CHiME3 simulated test set mentioned in Section IV.B. As performance scores, we used the short-time objective intelligibility measure (STOI) [28], in addition to the PESQ and SDR.

Table VI presents the scores of the initial and sixth SIBF outputs using the three source models, in addition to the initial reference. The scores of the initial SIBF output using the TV Gaussian model and the initial reference are identical to those in our previous study [8]. Similar to Table V, this table includes the scores of the sixth reference and Oracle SIBF using the TV Gaussian model. The optimal scores in each column are expressed in bold type, although the sixth references and Oracle SIBF were not considered for the comparison. As a common trend for all the source models, the scores of the sixth SIBF outputs were higher than those of the initial outputs, in addition to the initial reference, in terms of the three scores. Among the three source models, the BS Laplacian model demonstrated the most significant improvement by iterative casting.

The bottom four rows of Table VI present the scores of the observation using Microphone #5 and the methods that combine a conventional BF and DNN trained with the CHiME3 training dataset, with reference to the studies conducted by Heymann et al. [9], Erdogan et al. [11], and Tu et al. [12]. The scores of the method by Heymann et al. [9] are reported in [1] and subject to the speech distortions discussed in [9]. It should be noted that higher SDR does not always imply more accurate extraction, considering that this score can be sensitive to residual interferences in the lower frequencies as discussed in Section IV.E.2.

Heymann et al. and Erdogan et al. developed the mask-based Max SNR BF and minimum variance distortionless response (MVDR) BF, respectively. The SIBF outperformed these methods even in the initial casting, although there is a high probability that they were trained using the same dataset as that used in this study.

Tu et al. presented another variation of the mask-based MVDR BF, in which the mask is updated by feeding the BF output to the DNN, and the BF output is refined using a specific DNN for a postprocess. The PESQ score of the sixth SIBF output was identical to that of this method, despite not using such specific postprocesses. In addition, for the STOI score, the SIBF outperformed this method, even in the initial casting.

Table VI indicates that the sixth references outperformed the corresponding SIBF outputs for the three scores and the Oracle SIBF, except for the STOI. A discussion on the effective use of this feature is presented in the following section.

## V. FURTHER DISCUSSION ON BEHAVIORS OF THE SIBF

Focusing the relationship between the SIBF output and reference in terms of accuracy, we can briefly summarize the experimental results as follows:

1. In the initial casting, the SIBF output basically outperformed the reference.
2. After the second casting, however, the SIBF output basically underperformed the corresponding reference and outperformed the initial reference.

These two behaviors seem to be in opposition. However, we found that these can be understood consistently by introducing the concept of the fixed point. In this section, we first explain this concept. Thereafter, we discuss the further improvement of the SIBF performance based on the abovementioned analysis.

### A. What is Fixed Point?

Based on the experimental results, we classify the behaviors of the SIBF into the following two categories:

**Category 1**: the output is more accurate than the reference.

**Category 2**: the output is less accurate than the reference. Throughout all the results, we confirmed that the maximum of the outputs in Category 1 was lower than the minimum of those in Category 2 when using the same source model in the same scenario. For instance, for the BG × 2.0 scenario in Fig. 14, PESQ score of the initial SIBF output, which was the maximum in Category 1, was lower than that of the second SIBF output, which was the minimum in Category 2. These findings convinced us that, on the border between the two categories, there is a fixed point where the output is identical to the reference in terms of accuracy.

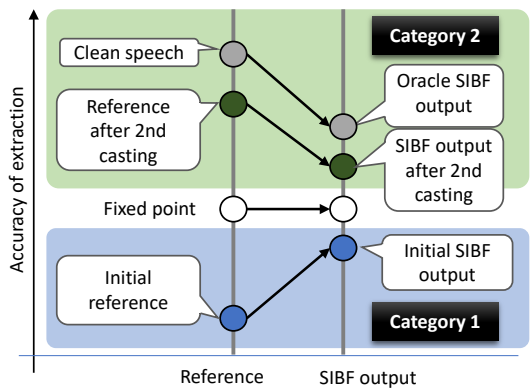

Fig. 16: Schematic plot of relationships between the references and SIBF outputs

The categories and fixed point are shown in Fig. 16, which presents a schematic plot of the relationship between the references and SIBF outputs. The vertical axis denotes the accuracy of the extraction, such as the PESQ and SDR. Category 1 includes the operation for generating the initial SIBF output in all the scenarios. Category 2 includes the operations after the second casting for all the scenarios, except that of BG × 0.25. Category 2 also includes the Oracle SIBF, given that its output is inevitably less accurate than the Oracle reference, namely clean speech. From the discussion in Section IV.E.2, we consider that Category 1 includes the operations in both the initial casting for the BG × 2.0 scenario and iterative casting for the BG × 0.25 scenario, despite SDR.

Similar to the accuracy of the SIBF output, the height of the fixed point depends on the SNR of the observation (namely, the scenario), type of source model, and the hyperparameters of each source model.

### B. Understanding Behaviors of SIBF by Using Fixed-Point

The concept of fixed point facilitates understanding the behaviors of the SIBF. This suggests that the generation of any SIBF output that is more accurate than the fixed point requires a reference that is more accurate than the fixed point. In the experiments, the initial references in all scenarios were less accurate than each corresponding fixed point. This indicates that unless the references are updated, the SIBF continues to behave as Category 1, and its outputs do not exceed the fixed point.

The results after the second casting suggest that iterative casting is a simple but effective method that can generate a reference more accurately than the fixed point even if the initial reference does not exceed the fixed point. In the results (except for the BG × 0.25 scenario), the reference exceeded it in the second casting, and the SIBF exhibited the behavior classified as Category 2.

However, to generate such references, the DNN output should be more accurate than the corresponding DNN output. If this condition is not satisfied, the SIBF exhibits the behavior classified as Category 1, and thus the SIBF outputs do not exceed the fixed point. Such phenomena occurred in the BG × 0.25 scenario.

### C. Techniques to Improve the Performance of the Iterative Casting

This subsection presents a discussion on techniques to improve the performance of iterative casting. A minimum of three independent methods can be considered in future studies.

An effective method in the initial casting involves the automatic adaptation of the hyperparameters of the source model to overcome sensitivities to the hyperparameters. For example, Fig. 5 suggests that adaptively selecting $\beta = 1/8$ or $\beta = 32$ instead of $\beta = 8$ can improve the SIBF using the TV Gaussian model in the BG × 2.0 scenario. It can improve the SIBF using other source models, because it can be employed as the boost start in the first iteration. Similarly, Fig. 7 suggests that adaptively selecting $\alpha = 0.01$ instead of $\alpha = 100$ can improve the SIBF using the BS Laplacian model in the same scenario.

An effective method in iterative casting involves the use of the reference after the second casting, instead of the SIBF output as the output of the extraction system. The concept of the fixed point suggests that if the reference exceeds the fixed point, it is more accurate than the corresponding SIBF output. Therefore, if we establish a method to determine whether the reference exceeds the fixed point, we can select a more accurate signal between the reference and previous SIBF outputs.

An alternative method involves the refining of the DNN to generate a more accurate reference by augmenting the training data. Generally, this is achieved by adding speech data with more varying SNRs. To improve the performance, particularly after the second casting, adding the SIBF outputs to the training data would be effective, considering that this can mitigate the mismatch between the DNN input and SIBF output.

## VI. RELATED WORKS

Considering that the SIBF is formulated as a deflationary ICA that utilizes the dependence between the reference and estimated target source, the SIBF is closely related to a series of studies on reference-based ICA. The SIBF exhibits another aspect of the beamformer combined with the DNN. A detailed examination of both aspects is presented in this section.

### A. Relationship between SIBF and Reference-based ICA

As an extension of the ICA, the SIBF has a minimum of two characteristics:

1. The combination of both the deflationary ICA and reference.
2. Separation, considering both independence and dependence.

A discussion of both characteristics is presented below.

#### 1) Combination of Both Deflationary ICA and Reference

The SIBF is based on the following concept: the combination of the deflationary ICA and reference allows for the extraction of only the source associated with the reference, without separating all the sources. The same concept can be found in studies conducted on the *ICA with reference* [35][36][37]. The difference between the SIBF and the schemes in the abovementioned studies is that the SIBF can combine a real-valued reference with complex-valued inputs and outputs, whereas the related studies handle real-valued signals.

#### 2) Separation Considering Both Independence and Dependence

The SIBF is based on a separation that utilizes both independence and dependence. This is related to the framework of the independent vector analysis (IVA) [20][21][38][39]. The IVA assumes that all the frequency bins in the same spectrogram are mutually dependent, whereas the spectrograms are mutually independent. The objective of assuming mutual dependence is to solve the frequency permutation problem that the order of the estimated sources is inconsistent among the frequency bins. To represent the dependence, the IVA uses a type of source model called an MS distribution, which can be expressed as (38).

$$p(y_k(1,t), \dots, y_k(F,t)) \propto g\left(\sqrt{\textstyle\sum_f |y_k(f,t)|^2}\right), \qquad (38)$$

where g(·) denotes an appropriate univariate PDF such as the univariate Laplacian distribution.

To improve the separation performance or intentionally control the order of the estimated sources, two different approaches that can reflect the supervisory information associated with the sources to the IVA have been examined. The first approach involves the use of the time-variant Gaussian model (39) to consider the time-varying variance as supervisory information [40].

$$p(y_k(1,t), \dots, y_k(F,t)|\sigma_k(t)) \\ \propto \exp\left(\frac{\sum_f |y_k(f,t)|^2}{\sigma_k(t)^2}\right), \qquad (39)$$

where $\sigma_k(t)^2$ denotes the time-varying variance estimated for the $k$th source. In [40], this variance was interpreted as a parameter of the source model. In terms of associating the estimated source with the reference, we can regard this model as an origin of the TV models in this study.

The second approach involves appending the time-varying variance to the MS model as another dependent component [41], as expressed by (40).

$$p(y_k(1,t), \dots, y_k(F,t), \sigma_k(t)) \\ \propto \exp\left(\sqrt{\textstyle\sum_f |y_k(f,t)|^2 + \gamma^2 \sigma_k(t)^2}\right), \qquad (40)$$

where $\gamma$ denotes the weight to control the influence of the variance. In terms of the aforementioned association, we can regard this model as an origin of the BS models in this study. This approach was extended to extract the utterance of the desired speaker [42]. The extension was derived combining (40) with the framework of extracting only the non-Gaussian source [43][44].

It should be noted that in these approaches, the time-varying variance $\sigma_k(t)^2$ was shared over all the frequency bins, and thus, it was qualitatively different from the reference used in this study.

Unlike the two approaches mentioned above, the SIBF uses a magnitude spectrogram as a reference that contains different data in each frequency bin. Moreover, the formulation of the SIBF assumes that all the frequency bins, even those in the same spectrogram, are mutually independent, as mentioned in Section II.B. This is because associating the estimated target source with the reference in each frequency bin can lead to the solution of the frequency permutation problem. A similar concept can be found in the model-based IVA [15] and IDLMA [16]. Both methods use time-frequency-varying variances, which are different in each frequency bin and frame. Such variances are generated using the spectral subtraction method in [15] or the DNN in [16].

For uniqueness in the formulation of the SIBF, we employ both TV and BS models and interpret that both types commonly represent the dependence between the reference and estimated target source, whereas the aforementioned related studies focus only on one type. Thanks to this interpretation, we found that the update rules for the BS Laplacian and TV t models are similar, as mentioned in Section II.C.3. These findings will contribute to the advancement in the reference-based ICA.

### B. Unified Formulation of Both SIBF and Mask-based Max SNR BF

Considering the formulation of the SIBF, this might seem to have few theoretical relationships to the conventional BFs. However, the SIBF is particularly related to the mask-based Max SNR BF [9][10] because both BFs can be formulated in a unified manner as detailed below.

In this subsection, $\boldsymbol{v}_1(f)$, instead of $\boldsymbol{w}_1(f)$, is used as the extraction filter that directly extracts the target source from the observations, as expressed by (41).

$$y_1(f,t) = \boldsymbol{v}_1(f)^{\mathrm{H}} \boldsymbol{x}(f,t). \tag{41}$$

As mentioned below, we found that both BFs are commonly formulated as the minimization problem using a weighted covariance matrix $\boldsymbol{C}(f)$ expressed as (42) and (43) under Constraint (44), and that the differences between both BFs are represented by calculating the weight $c(f,t)$.

$$\boldsymbol{C}(f) = \langle c(f,t)\boldsymbol{x}(f,t)\boldsymbol{x}(f,t)^{\mathrm{H}} \rangle_t, \tag{42}$$

$$\boldsymbol{v}_1(f) = \arg\min_{\boldsymbol{v}_1(f)} \boldsymbol{v}_1(f)^{\mathrm{H}} \boldsymbol{C}(f)\boldsymbol{v}_1(f), \tag{43}$$

$$\text{s.t. } \boldsymbol{v}_1(f)^{\mathrm{H}} \boldsymbol{\Phi}_x(f)\boldsymbol{v}_1(f) = 1, \tag{44}$$

where $\boldsymbol{\Phi}_x(f)$ is the observation covariance matrix defined in (3). The filter $\boldsymbol{v}_1(f)$ can be obtained as the solution of the following generalized eigenvalue problem [19]:

$$\boldsymbol{C}(f)\boldsymbol{v}_1(f) = \lambda_{\min}(f)\boldsymbol{\Phi}_x(f)\boldsymbol{v}_1(f), \tag{45}$$

where $\lambda_{\min}(f)$ is the minimum eigenvector. We express the rule for $\boldsymbol{v}_1(f)$ as follows:

$$\boldsymbol{v}_1(f) = \mathrm{GEV}_{\min}(\boldsymbol{C}(f), \boldsymbol{\Phi}_x(f)), \tag{46}$$

where $\mathrm{GEV}_{\min}$ denotes the generalized eigenvector corresponding to the minimum eigenvalue of the given two Hermitian matrices.

Thereafter, we transform the update rules of the SIBF to match them to (43), whereas the rule of the mask-based Max SNR BF is transformed as shown in the Appendix.

First, we transform (15), which is the minimization problem for $\boldsymbol{w}_1(f)$, to that for $\boldsymbol{v}_1(f)$. From (7) and (41), $\boldsymbol{w}_1(f)$ can be expressed as follows:

$$\boldsymbol{w}_1(f) = (\boldsymbol{P}(f)^{\mathrm{H}})^{-1}\boldsymbol{v}_1(f). \tag{47}$$

Applying (47) to both (15) and (16) leads to the following minimization problem:

$$\boldsymbol{v}_1(f) = \arg\min_{\boldsymbol{v}_1(f)} \{-\langle \log \mathrm{p}\left(r(f,t), y_1(f,t)\right)\rangle_t\} \tag{48}$$

$$\text{s.t. } \boldsymbol{v}_1(f)^{\mathrm{H}} \boldsymbol{\Phi}_x(f)\boldsymbol{v}_1(f) = 1. \tag{49}$$

Constraint (49) is equivalent to $\langle |y_1(f,t)|^2\rangle_t = 1$.

To derive the rule for the TV Gaussian model, we apply (17) to (48) and then obtain the following minimization problem under Constraint (49).

$$\boldsymbol{v}_1(f) = \arg\min_{\boldsymbol{v}_1(f)} \left\{ \boldsymbol{v}_1(f)\langle \frac{\boldsymbol{x}(f,t)\boldsymbol{x}(f,t)^{\mathrm{H}}}{r(f,t)^{\beta}}\rangle_t \boldsymbol{v}_1(f)^{\mathrm{H}} \right\}. \tag{50}$$

To prevent division by zero, we replace $r(f,t)$ in (50) with $r'(f,t)$ defined in (20). Consequently, we obtain the following correspondence between (43) and (50):

$$c(f,t) = \min\left(1, \ \frac{\varepsilon}{r(f,t)^{\beta}}\right), \tag{51}$$

where $\min(\cdot)$ denotes the minimum argument and $\varepsilon$ is the clipping threshold used in (20). In deriving (51), we multiplied $\varepsilon$ by the right-hand side of (50) because it does not change the solution in (50).

Similarly, for the BS Laplacian and TV t models, we obtain the correspondences expressed as (52) and (53), respectively, through applying (22) and (29) to (48).

$$c(f,t) = \min\left(1, \ \frac{\varepsilon}{\sqrt{\alpha r(f,t)^2 + |y_1(f,t)|^2}}\right), \tag{52}$$

$$c(f,t) = \min\left(1, \ \frac{(\nu + 2)\varepsilon}{\nu r(f,t)^2 + 2|y_1(f,t)|^2}\right), \tag{53}$$

In estimating $\boldsymbol{v}_1(f)$ for both models, a sequence of iterative rules is used. For the BS Laplacian model, this sequence consists of (41), (52), (42), and (46). For the TV t model, (53) is used instead of (52).

As mentioned in the Appendix, for the mask-based Max SNR BF, the mask for the interferences corresponds to the weight $c(f, t)$. By analogy with the SIBF, however, the weight can also be interpreted as follows:

$$c(f, t) = \begin{cases} 1, & \text{(Target is not present.)} \\ 0, & \text{(Otherwise)} \end{cases} \tag{54}$$

We can compare both BFs because these are represented in a unified form. The comparison results suggest that the SIBF uses a particular mask consisting of continuous values. Specifically, the mask uses values close to 0, such as $\varepsilon / r(f, t)^\beta$, in (51), (52), and (53) instead of 0 in (54). The use of 1 in these equations corresponds to clipping to $\varepsilon$ in (20), (27), and (33) respectively, and an ascend in the clipping threshold $\varepsilon$ increases the rate of 1 in the mask.

The unified formulation is particularly useful for the SIBF using the TV Gaussian model, because it can clarify the effects of the reference exponent $\beta$ in (20). From an examination of (51), an increase in $\beta$ has a minimum of two effects:

a)  An increase in $\beta$ increases the rate of 1 in the mask, given that clipping to 1 occurs when $r(f, t) < \varepsilon^{1/\beta}$ in (51).

b)  An increase in $\beta$ makes the mask more similar to the binary mask, given that $\varepsilon / r(f, t)^\beta$ approaches 0. The case of $\beta \to \infty$ is equivalent to (54).

These two effects separately demonstrate the peak performances for different values of $\beta$ and can explain the two typical peaks in the results of the TV Gaussian model, as shown in Fig. 5 and Fig. 6.

## VII. Conclusions

In this paper, we detailed a method for target source extraction, referred to as SIBF, and its advancements. The SIBF employs a rough magnitude spectrogram of the target source as the initial reference to extract the target source more accurately. For the extraction, we extended the framework of deflationary ICA by considering the dependence between the reference and extracted target source, in addition to the mutual independence of all potential sources. To solve the extraction problem using ML estimation, we examined three source models that can represent the dependence, namely, the TV Gaussian, BS Laplacian, and TV t models. For each model, we derived the update rules for the extraction filter.

To improve the performance, we developed two methods, as follows. The first was the boost start, which used the optimal configuration for the TV Gaussian model in the first iteration to accelerate the convergence of the SIBF using the other models. The second was iterative casting, which casted the previous SIBF output into the reference-generating method to obtain a more accurate reference than the initial reference.

The advantage of the SIBF is that it can generate a more accurate signal than the spectrogram generated by target-enhancing methods, such as DNN-based speech enhancement. We verified this through experiments using the CHiME3 dataset and confirmed that the boost start and iterative casting further enhanced this feature. This is the first advancement from the previous study.

Based on the experimental results, the concept of the fixed point was developed, which suggested that the reference should exceed the fixed point for the SIBF outputs to be more accurate than the fixed point. We confirmed that iterative casting is a simple and effective method to generate such a reference when the initial reference does not exceed the fixed point. This concept facilitated both understanding the behaviors of the SIBF and discussing further improvement of the extraction performance. This is the second advancement.

The third advancement is the unified formulation of the SIBF and mask-based Max SNR BF, which were formulated as a minimization problem using a weighted covariance matrix, and the calculation of the weight characterized each BF. This formulation allows for the knowledge of the mask-based Max SNR BF to be applied to the SIBF.

Future work will focus on the following aspects. The first is further verification of the effectiveness of the SIBF by using more varying performance scores and tasks. We consider the use of the word error rate in ASR as a performance score and the task of extracting a single speech source from overlapping speech data by employing the DNN trained for this task. The second is the further improvement of the extraction accuracy, as discussed in Section V.C. We consider that developing a novel source model that involves the three models is also significant for the improvement, considering that the update rules for both the BS Laplacian and TV t models seem to be similar, as mentioned in Section II.C.2 and II.C.3. The third is the introduction of various techniques used in conventional BFs and the ICA to the SIBF.

Finally, this study also contributes to the ICA and BF in the following aspects:

1.  The analysis of the TV Gaussian and TV t models contributes to the studies that employ these models, such as [15], [16], and [26].

2.  The concept of the fixed point contributes to the studies that use the DNN outputs as the time-frequency-varying variances, such as [16], considering that these variances can be evaluated as a sound by applying the ISTFT.

3.  The unified formulation contributes to mutual use of the expertise between the ICA and BFs, considering that the mask-based BFs can be interpreted as a particular case of deflationary ICA. For example, the theories for ICA can be applied for developing a novel BF, as we did in this study, whereas the technologies on the mask-based BF, including training of DNNs, can be applied for developing a novel variation of DNN-based ICA.

We hope that this study promotes further research in both fields.

## Appendix

We transform the rule of the mask-based Max SNR BF [9][10] to match it to (43), which represents the constrained minimization problem using a weighted covariance matrix.

The formulation of the mask-based Max SNR BF is first presented. If $m_T(f, t)$ and $m_I(f, t)$ are binary masks for the target source and interferences, respectively, they are defined as follows:

$$m_T(f, t) = \begin{cases} 1, & \text{(Only the target is present)} \\ 0, & \text{(Otherwise)} \end{cases} \quad (55)$$

$$m_I(f, t) = \begin{cases} 1, & \text{(Only interferences are present)} \\ 0, & \text{(Otherwise)} \end{cases} \quad (56)$$

It should be noted that in [9] and [10], both masks were estimated by the DNN and consisted of continuous values because no binarization processes were applied. However, we treat these masks as binary ones, considering that the DNN itself was trained with the ideal binary masks.

The Max SNR BF is formulated as the solution of the following maximization problem:

$$\boldsymbol{\Phi}_T(f) = \frac{\sum_t m_T(f, t) \boldsymbol{x}(f, t) \boldsymbol{x}(f, t)^H}{\sum_t m_T(f, t)}, \quad (57)$$

$$\boldsymbol{\Phi}_I(f) = \frac{\sum_t m_I(f, t) \boldsymbol{x}(f, t) \boldsymbol{x}(f, t)^H}{\sum_t m_I(f, t)}, \quad (58)$$

$$\boldsymbol{v}_1(f) = \arg\max_{\boldsymbol{v}_1(f)} \frac{\boldsymbol{v}_1(f)^H \boldsymbol{\Phi}_T(f) \boldsymbol{v}_1(f)}{\boldsymbol{v}_1(f)^H \boldsymbol{\Phi}_I(f) \boldsymbol{v}_1(f)}, \quad (59)$$

where $\boldsymbol{\Phi}_T(f)$ and $\boldsymbol{\Phi}_I(f)$ are the target and interference covariance matrices, respectively. The filter $\boldsymbol{v}_1(f)$ can be obtained as the solution of the following generalized eigenvalue problem.

$$\boldsymbol{\Phi}_T(f) \boldsymbol{v}_1(f) = \lambda_{max}(f) \boldsymbol{\Phi}_I(f) \boldsymbol{v}_1(f), \quad (60)$$

where $\lambda_{max}(f)$ denotes the maximum eigenvector.

Thereafter, to make (59) comparable to (43), we exclude $\boldsymbol{\Phi}_T(f)$ and rewrite (59) as the constrained minimization problem.

By assuming that all the sources are stationary during the segment for the covariance calculation, such as (57) and (58), the target covariance $\boldsymbol{\Phi}_T(f)$ can be substituted as follows [45]:

$$\boldsymbol{\Phi}_T(f) = \boldsymbol{\Phi}_x(f) - \boldsymbol{\Phi}_I(f), \quad (61)$$

where $\boldsymbol{\Phi}_x(f)$ is the observation covariance matrix defined in (3). Given that the solution in (59) does not change even if the right-hand side is multiplied by a scalar value, we can use $\boldsymbol{\Phi}_I'(f)$ instead of $\boldsymbol{\Phi}_I(f)$:

$$\boldsymbol{\Phi}_I'(f) = \langle m_I(f, t) \boldsymbol{x}(f, t) \boldsymbol{x}(f, t)^H \rangle_t, \quad (62)$$

From both (61) and (62), we can express (59) as (63).

$$\boldsymbol{v}_1(f) = \arg\max_{\boldsymbol{v}_1(f)} \frac{\boldsymbol{v}_1(f)^H \boldsymbol{\Phi}_x(f) \boldsymbol{v}_1(f)}{\boldsymbol{v}_1(f)^H \boldsymbol{\Phi}_I'(f) \boldsymbol{v}_1(f)}, \quad (63)$$

The maximization problem (63) is equivalent to the minimization problem (64), considering that both the numerator and denominator in (63) are always positive.

$$\boldsymbol{v}_1(f) = \arg\min_{\boldsymbol{v}_1(f)} \frac{\boldsymbol{v}_1(f)^H \boldsymbol{\Phi}_I'(f) \boldsymbol{v}_1(f)}{\boldsymbol{v}_1(f)^H \boldsymbol{\Phi}_x(f) \boldsymbol{v}_1(f)}. \quad (64)$$

The minimization problem (64) can be transformed to the constrained minimization problem expressed as (65) and (44), considering that (59) can be transformed to the corresponding constrained maximization problem [45].

$$\boldsymbol{v}_1(f) = \arg\min_{\boldsymbol{v}_1(f)} \{ \boldsymbol{v}_1(f)^H \boldsymbol{\Phi}_I'(f) \boldsymbol{v}_1(f) \}. \quad (65)$$

Comparing (65) with (43), we obtain the simple correspondence that $c(f, t) = m_I(f, t)$.

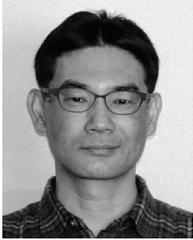

**Atsuo Hiroe** (M'11) received B.S. and M.S. degrees in computer science from Tokyo Institute of Technology (Tokyo, Japan) in 1994 and 1996, respectively.

From 1996–2014, he worked at Sony Corporation (Tokyo, Japan), conducting research and development on speech recognition, speech signal processing, and natural language understanding, among other fields. From 2014–2016, he was seconded to the National Institute of Information and Communications Technology (Kyoto, Japan) to study spoken dialog systems. From 2016 onward, he has worked at Sony Group Corporation, conducting research and development on speech signal processing.